\documentclass{aa}
\usepackage{graphicx}
\usepackage{txfonts}
\usepackage{orcidlink}
\usepackage{lastpage}
\usepackage{booktabs}
\usepackage{cleveref}
\newcommand{\icarogw}{\textsc{icarogw}\xspace}

\newcommand{\Msol}{\xspace\rm{M}_{\odot}\xspace}

\newcommand{\snr}{\textsc{SNR}\xspace} 
\newcommand{\de}{{\rm d}}

\usepackage{enumitem,amssymb}
\newlist{todolist}{itemize}{2}
\setlist[todolist]{label=$\square$}
\usepackage{pifont}

\begin{document} 

   \title{Inferring astrophysics and cosmology with individual compact binary coalescences and their gravitational-wave 
stochastic background}
   \author{S.~Ferraiuolo
          \inst{1,2,3}
          \orcidlink{0009-0005-5582-2989} \thanks{sarah.ferraiuolo@univ-amu.fr},
          S. Mastrogiovanni
          \inst{2}
          \orcidlink{0000-0003-1606-4183},
          S.~Escoffier\inst{3}
          \orcidlink{-} \and
          E.~Kajfasz\inst{3}
          \orcidlink{-}         
          }
   \institute{Universit\'a di Roma La Sapienza, 00185 Roma, Italy
         \and
            INFN, Sezione di Roma, I-00185 Roma, Italy
         \and 
             Aix Marseille Univ, CNRS/IN2P3, CPPM, Marseille, France
             }
  \abstract
{Gravitational waves (GWs) from compact binary coalescences (CBCs) provide a new avenue to probe the cosmic expansion, in particular the Hubble constant $H_0$. The spectral sirens method is one of the most used techniques for GW cosmology. It consists of obtaining cosmological information from the GW luminosity distance, directly inferred from data, and the redshift that can be implicitly obtained from the source frame mass distribution of the CBC population. With GW detectors, populations of CBCs can be either observed as resolved individual sources or implicitly as a stochastic gravitational-wave background (SGWB) from the unresolved ones. In this manuscript, we study how resolved and unresolved sources of CBCs can be employed in the spectral siren framework to constrain cosmic expansion. The idea stems from the fact that the SGWB can constrain additional population properties of the CBCs thus potentially improving the measurement precision of the cosmic expansion parameters.
We show that with a five-detector network at O5-designed sensitivity, the inclusion of the SGWB will improve our ability to exclude low values of $H_0$ and the dark matter energy fraction $\Omega_m$, while also improving the determination of a possible CBC peak in redshift. Although low values of $H_0$ and $\Omega_m$ will be better constrained, we obtain that most on the precsion on $H_0$ will be provided by resolved spectral sirens.

We also performed a spectral siren analysis for 59 resolved binary black hole sources detected during the third observing run with an inverse false alarm rate higher than 1 per year jointly with the SGWB. We obtain that with current sensitivities, the cosmological and population results are not impacted by the inclusion of the SGWB. 
}


\maketitle

\section{Introduction}
\label{sec:introduction}
% Your introduction here
Gravitational waves (GWs) from compact binary coalescences (CBCs) offer a unique opportunity to provide new insight into the study of the expansion of the Universe, especially in light of ongoing tensions in cosmology (\cite{Di_Valentino_2021}). Unlike electromagnetic (EM) observables, CBCs detected through GWs carry direct information about their luminosity distance, as such they are referred to as \textit{standard sirens} (\cite{Schutz_1986,2005ApJ...629...15H}). When combined with redshift information, we can therefore use GWs data to probe the luminosity distance-redshift relation - and hence key cosmological parameters such as the Hubble constant $H_0$. However, GWs alone do not provide the source redshift. Various methods have been developed to estimate redshifts for GW events, which are essential for this purpose.

\textit{Bright sirens} are a special class of standard sirens, where an EM counterpart is observed along with the GW emission. This was the case of the binary neutron star merger GW170817 detection in association with a gamma-ray burst and a subsequent kilonova (\cite{GW170817}). The host galaxy for this source was identified, allowing us to obtain a precise redshift measurement and an $\textrm{H}_0$ estimate of $70^{+19}_{-8}$ km/s/Mpc (\cite{GW_standard_sirens}). So far, GW170817 is the only event to date with an associated EM counterpart. 

Most often, other detections involve \textit{dark sirens}, such as binary black hole (BBH) mergers, with no EM counterparts. For these types of events, several methods have been proposed to obtain a statistical redshift estimation for the source. One approach involves cross-matching the event’s sky localization with galaxy catalogs, assigning probabilities based on the distribution of potential host galaxies \citep{LIGOScientific:2017ycc,DES:2019,LIGOScientific:2019zcs,gray2020,palmese20_sts,Finke_2021,Gray_2022,Mukherjee:2022afz,GWTC3_expansion,Mastrogiovanni2023,2023JCAP...12..023G,darksiren_DESI,Bom24_darksiren}. Another method, clustering-based techniques exploit correlations between gravitational-wave events and large-scale structure to constrain redshifts \citep{Namikawa:2015prh,Oguri:2016dgk,Namikawa:2016edr,Zhang:2018nea,Scelfo:2020jyw,Bera:2020jhx,Libanore:2020fim,Mukherjee:2020hyn,diaz22,Ferri:2024amc,Ghosh:2023ksl,Zazzera:2024agl}.

In this paper we focus on the \textit{spectral sirens} method \citep{Taylor2012,Farr_2019,2021PhRvD.104f2009M,Mancarella_2022,Ezquiaga:2020tns,spectral_sirens,Mali:2024wpq,Mancarella:2024qle}.  
This approach uses the inherent degeneracy between a source's redshift $z$ and its detector frame mass $m_{\rm det}$, as well as its source frame mass $m_s$, linked by the relation $m_{\rm det}=(1+z)m_s$. The sources' redshifts are indirectly inferred by measuring the detector frame masses and simultaneously modeling the source mass distribution and the cosmological parameters. The effectiveness of this method is influenced by the choice of phenomenological models used to represent the CBC source mass distribution and their ability to accurately capture the true characteristics of the CBC population \citep{Mukherjee_2022, Gray_2022, Karathanasis:2022rtr, Pierra:2023deu, Mastrogiovanni_2023,Agarwal:2024hld}. This approach was applied to measure $H_0$ using 47 gravitational-wave sources from the Third LIGO and Virgo Gravitational-Wave Transient Catalog (GWTC–3)  obtaining $H_0=68^{+12}_{-7}$ km/s/Mpc  (68\% credible interval) when combined with the $H_0$ measurement from GW170817 and its electromagnetic counterpart (\cite{GWTC3_cosmo}).

In addition to resolved sources, we expect to detect a stochastic gravitational-wave background (SGWB) from unresolved sources. The SGWB depends on the properties of the population and the underlying population of CBCs (\cite{Romano_2017,PhysRevD.46.5250, 2023PrPNP.12804003V}). The idea is to explore whether the detection or non-detection of this stochastic background, combined with the information from resolved sources, can help constrain the population properties more accurately and, in turn, refine the measurements of the cosmological parameters \citep{SGWB_LVK}. Moreover, as recognized in \citet{cousins_2025}, the SGWB also has a direct dependency from $H_0$ given by the fact that the GW energy density is related to the comoving volume. The interplay between GW signals from CBCs and the SGWB presents an exciting opportunity to improve our understanding of fundamental cosmological and astrophysical parameters.

The method of synthesizing data from individual CBC events with upper limits on the SGWB was initially examined by \cite{Callister_2020}, \cite{Turbang_2024} in the context of astrophysical properties for CBCs and more recently by \citet{cousins_2025} for a joint cosmological and population analysis with the latest Gravitational-Wave Transient Catalog (GWTC-3). In our study, we focus more on understanding what population properties and cosmological parameters could be constrained by the SGWB for future observing runs.

This paper is organized as follows. Section \ref{sec:Analysis Method} outlines the methodology for our integrated analysis of individual binary black hole mergers and the stochastic background. Section \ref{sec:Data} presents how we simulate the data. We discuss our projection studies in Sec.~\ref{sec:Result} and a reanalysis with this methodology of GWTC-3 in Sec.~\ref{sec:Result_data}. Finally in Sec.~\ref{sec:conc} we summarize our conclusions.

\section{Analysis Method}
\label{sec:Analysis Method}

We aim to infer the cosmological and population parameters including resolved and unresolved CBC. We build on the framework established by \cite{Callister_2020} and \cite{Turbang_2024}. We describe the GW dataset as a set $\{x\}$ of $N_{ \rm obs}$ resolved GW detections found in data chunks $\{x\}= \{x_1, ..., x_{N_{\rm obs}}\}$ and a set of elements $\hat{C}(f)$ describing the amount of correlated noise across GW detectors that are obtained for an observing time $T_{\rm obs}$. Let $\Phi$ denote the hyperparameters that characterize the CBC population and cosmology (e.g., $H_0$ and $\Omega_m$). We proceed under the assumption that the joint likelihood of these observations can be factorized as:
\begin{equation}
    \mathcal{L}(\{x\} , \hat{C}|\Phi) = \mathcal{L}_{\rm CBC} (\{x\} |\Phi) \mathcal{L}_{\rm SGWB}(\hat{C}|\Phi),
    \label{eq:combined likelihood}
\end{equation}
where $\mathcal{L}_{\rm CBC}$ represents the likelihood associated with the detection of resolved sources, which, as discussed in Sec.~\ref{subsec:Hierarchical Bayesian Likelihood}, follows an inhomogeneous Poisson distribution. The likelihood $\mathcal{L}_{\rm SGWB}$ denotes stochastic likelihood modelled as a Gaussian process, as detailed in Sec.~\ref{subsec:SGWB Likelihood}. 
We note that the factorization of the likelihood in Eq.~\ref{eq:combined estimator} assumes that the ``stochastic'' data is independent of the resolved sources. In other words, we assume that the amount of stochastic GW signal is not modified after removing all the resolved sources in the observing period $T_{\rm obs}$. This might not be the case when the sensitivity of the GW detectors is enough to solve a considerable amount of the population. To address this issue while still using the same likelihood model, we propose splitting the GW data collected in an observing time $T_{\rm obs}$ into two disjoint sets with duration $T_{\rm obs}/2$ and use one for detecting solved sources while the other for the SGWB.

\subsection{Hierarchical Bayesian Likelihood}
\label{subsec:Hierarchical Bayesian Likelihood}
The detection of GWs events is modeled as an inhomogeneous Poisson process with selection biases (\cite{10.1093/mnras/stz896,Vitale_2021}). For $N_{obs}$ detected GWs signals over an observation time $T_{\textrm{obs}}$, the probability of obtaining a specific GWs dataset $\{x\}$ given population hyperparameters $\Phi$ is:
\begin{align}
    \mathcal{L}_{\text{CBC}}(\{x\}|\Phi) \propto  e^{- N_{\textrm{exp}}(\Phi)}&\prod^{N_{\textrm{obs}}}_{i=1}  T_{\textrm{obs}}
    \int d\theta dz \times \nonumber\\ 
    &\times \mathcal{L}_{\textrm{GW}}(x_i|\theta,z,\Phi)\frac{1}{1+z}\frac{dN_{\textrm{CBC}}}{d\theta dzdt_s}(\Phi),
    \label{eq: hierarchical likelihood}
\end{align}
here, $N_{\text{exp}}$ represents the expected number of  CBC detections within the observation time $T_{\text{obs}}$. Typically $N_{\rm exp}$ is evaluated using a set of Monte Carlo injections in real data and it is used to correct for selection biases, more details are provided in the App.~\ref{appendix A}. 
The variables $z$ and $\theta$ denote the redshift and a set of binary parameters for each CBC, such as the source masses. The individual GW likelihood $\mathcal{L}_{\text{GW}}(x_i |\theta, z, \Phi)$ incorporates uncertainties in the parameters $\theta$; the final term of the equation corresponds to the CBC rate in the source frame time $t_{\rm s}$. The model that we adopt for the CBC rate will be described in more detail in Sec.~\ref{sec:Data}.

\subsection{Stochastic Gravitational-Wave Background Likelihood}
\label{subsec:SGWB Likelihood}
 The SGWB is conventionally defined as \citep{Romano_2017}:
\begin{equation}
    \Omega_{\text{GW}}(f)=\frac{1}{\rho_c}\frac{\rm d\ln \rho_\text{GW}(f)}{{\rm d}\ln f},
\end{equation}
where $\rho_c=3H_0^2 c^2/(8\pi G)$ is the critical energy density of the Universe, $\rho_{\rm GW}(f)$ is the SGWB energy density observed at the frequency $f$, $G$ is Newton’s gravitational constant, $c$ is the speed of light, and $H_0$  is the Hubble constant.
The SGWB energy density modeled from the population of CBC is given by \citep{phinney2001practicaltheoremgravitationalwave}:
\begin{equation}
        \Omega_{\textrm{GW}}(f) = \frac{f}{\rho_{\rm c}} \int_0^{+\infty} dz \, \frac{ R(z)}{ H(z)(1+z)} \left\langle \frac{dE_{\rm s}}{df_{\rm s}} \bigg|_{f(1+ z)} \right\rangle,
      \label{eq: omega_gw}
\end{equation}
here, $R(z)$ is the event rate per unit comoving volume and per unit time at redshift $z$. The term $H(z)$ is the Hubble parameter that under the assumption of $\Lambda$CDM is defined as $H(z)=H_0\sqrt{\Omega_{\rm m}(1+z)^3+\Omega_\Lambda}$ \citep{Wei_2017}. As recognized in \citet{cousins_2025}, the SGWB scales as $H_0^{-3}$, this scaling arises from comoving volume density term ($\propto H_0^{-3}$).
The final term describes the energy spectrum radiated by the CBC population, averaged over the source population at a given redshift. More details on how we calculate the SGWB are provided in the App.~\ref{appendix A}

Searches for the SGWB, such as those conducted by the LIGO-Virgo-Kagra (LVK) collaboration (\cite{SGWB_LVK}), aim to measure the dimensionless energy density $\Omega_{\rm GW}(f)$ in Eq. \ref{eq: omega_gw}. Let us label the GW detectors in the network by the index $I$, the optimal cross-correlation estimator for a baseline $I J$ is given by \citet{Allen_1999,PhysRevD.104.022004}:
\begin{equation} 
\hat{C}_{IJ}(f) = \frac{2}{ T_{\rm obs}} \frac{10\pi^2}{3H_0^2} \frac{f^3}{\gamma_{IJ}(f)} \tilde{s}_I(f) \tilde{s}^*_J(f), 
\label{eq:cross-corr estimator general}
\end{equation}
where $T_{\rm obs}$ is the observation time, and $\tilde{s}_I(f)$ and $\tilde{s}_J(f)$ are the Fourier transforms of the data for detectors $I$ and $J $. The overlap reduction function $\gamma_{IJ}(f)$ encodes the baseline geometry of the detector pair $IJ$ (\cite{PhysRevD.46.5250,PhysRevD.48.2389}).
The combined cross-correlation estimator using information from all baselines is obtained by:
\begin{equation}
\hat{C}(f) = \frac{\sum_{IJ} \hat{C}_{IJ}(f)\sigma^{-2}_{IJ}}{\sum_{IJ}\sigma^{-2}_{IJ}},
\label{eq:combined estimator}
\end{equation}
with an expected value of $\langle \hat{C}(f) \rangle = \Omega_{\rm GW}(f)$. We adopt the shorthand notation $\sum_{IJ}$ meaning a sum over all independent baselines $IJ$.
The variance of the estimator for a single baseline $I J$ is:
\begin{equation}
\sigma_{IJ}^2(f) \approx \frac{1}{2T_{\rm obs}\Delta f}\left(\frac{10\pi^2}{3H_0^2}\right)^2\frac{f^6}{\gamma_{IJ}^2(f)}P_I(f)P_J(f),
\label{eq:variance single pair}
\end{equation}
under the assumption of a small signal-to-noise ratio for the SGWB, where $P_I(f)$ and $P_J(f)$ are the one-sided power spectral densities of detectors $I$ and $J$, and $\Delta f$ is the frequency resolution. The combined variance for all pairs, assuming statistical independence, is given by:
\begin{equation}
\sigma^2(f) = \left(\sum_{I J} \frac{1}{\sigma_{IJ}^2(f)}\right)^{-1}.
\label{eq:combined variance}
\end{equation}
The variance $\sigma^2(f)$ can be used to set the upper limit on the SGWB.

The cross-correlation estimator in Eq. \ref{eq:combined estimator} is then used in the likelihood $\mathcal{L}_{\text{SGWB}}$ , which is well-approximated by a Gaussian distribution (\cite{SGWB_LVK}):

\begin{equation}
    \mathcal{L}_{\text{SGWB}}(\hat{C}|\Phi) \propto 
    \exp\left[-\frac{1}{2}\sum_k  \left(\frac{\hat{C}(f_k)-\Omega_{\rm GW}(f_k,\Phi)}{\sigma(f_k)}\right)^2 \right], 
    \label{eq: stochastic likelihood}
\end{equation}
where the sum runs over discrete frequency bins $f_k$ and where $\Omega_{\rm GW}$ is our model energy-density spectrum defined in Eq. \ref{eq: omega_gw}. We note that the stochastic likelihood is not invariant for $H_0$, but it scales as $H_0^{-1}$. This scaling is due to the combination of the CBC energy density with the cross-correlation statistic and variance, which also matches the $1/H_0$ scaling of the SGWB SNR for detection \citep{cousins_2025}.

\section{Framework to simulate GW data for O5 sensitivities}
\label{sec:Data}

We build two observing scenarios for the simulation that are based on two years of GW data. The first one is called ``Resolved CBCs'' and it consists of 2 years of GW data that is used \textit{only} to detect resolved sources. The second case is called ``combined SGWB and CBCs'' and it is generated with 1 year of GW data used to collect resolved sources and 1 year used to detect the SGWB. The choice of dividing the GW data into two independent sets for the resolvable sources and the SGWB is driven by the fact that we want to keep the hierarchical likelihood separable. We also have a third validation case with 1 year of GW data employed only for stochastic searches that is used to understand the actual contribution of the SGWB on the population parameters.

The common ingredient needed to simulate data for the solved GW sources and the SGWB is a population and cosmological model for the CBCs. We assume a $\Lambda$CDM cosmology \citep{Wei_2017} and we parametrize the rate of CBCs introduced in Sec.~\ref{sec:Analysis Method} as:
\begin{align}
    \frac{\de  N_{\textrm{CBC}}}{\de\theta \de z \de t_s}= & R_0\psi(z,\Phi)p_{\text{pop}}(\Vec{m}_{\rm s}|\Phi)\frac{\de V_{\rm c}}{\de z}
\end{align}
where $R_0=20$ Gpc$^{-3}$yr$^{-1}$ is the local merger rate at $z=0$, $\psi(z, \Phi)$ is the merger rate evolution with redshift, and $ p_{\rm pop}(\Vec{m}_{\rm s}|\Phi)$ represents the probability density function for the source frame masses $\Vec{m}_{\rm s}=(m_{\rm 1,s}, m_{\rm 2,s})$. The term $\de V_{\rm c}/\de z$ is the differential of the comoving volume per unit redshift. For the source frame masses $m_{\rm 1,s}$ and $m_{\rm 2,s}$, we adopt a power-law plus peak (PLP) model, consistent with the LIGO and Virgo GWTC-3 results (\cite{PhysRevX.13.011048}): 
\begin{align}
    & p(m_{\rm 1,s} | m_{\text{min}}, m_{\text{max}}, \alpha) = (1 - \lambda_\text{g}) P(m_{\rm 1,s} | m_{\text{min}}, m_{\text{max}},\alpha) +\\
    & + \lambda_\text{g} G(m_{\rm 1,s} | \mu_{\rm g}, \sigma_{\rm g}), \nonumber \quad \text{with } (0 \leq \lambda_\text{g} \leq 1) \nonumber \\
    &p(m_{\rm 2,s} | m_{\text{min}}, m_{\rm 1,s}, \beta) = P(m_{\rm 2,s} | m_{\text{min}}, m_{\rm 1,s}, \beta),
    \label{eq:mass model}
\end{align}
where the primary mass $m_{\rm 1,s}$ follows a combination of a truncated power-law distribution and a Gaussian peak, weighted by the fraction $\lambda_{\text{g}}$. The secondary mass $m_{\rm 2,s}$ is modeled conditionally on $m_{\rm 1,s}$ using a power-law distribution. We use this model to describe a population of BBHs. The PLP model is motivated by the fact there is a potential accumulation of CBC around $35 M_\odot$ just before the pair-instability supernovae gap \citep{Talbot_2018}. In both distributions, we applied a smoothing window at the low end of the mass distribution dependent on a parameter $\delta_m$ representing the smoothing length. This is introduced to model the effects of the stellar progenitor metallicity \citep{Abbott_2021}.
The redshift evolution $\psi(z,\Phi)$, is modeled using a rate model \citet{madau_rate, Callister_2020}, which is expressed as:
\begin{equation}
     \psi(z, \Phi) = \left[1 + (1 + z_{\rm p})^{-\gamma - \kappa}\right] \frac{(1 + z)^{\gamma}}{\left(1 + \frac{1 + z}{1 + z_{\rm p}}\right)^{\gamma + \kappa}} ,
    \label{eq: madau rate}
\end{equation}
where $\gamma$ governs growth at low redshift, $z_{\rm p}$ marks the transition redshift, and $\kappa$ determines the decline at high redshift. The hyperparameters' values assumed for the population are reported in Tab.~\ref{tab:priors}. According to the rate model, the total number of BBH coalescences in one year in all of the universe is about $10^6$, which agrees with existing literature \citep{Borhanian_2024,Ng_2021}. 

\subsection{Resolved Gravitational Wave Sources}

To simulate resolved GW sources, we generate a population of BBHs following the previously described population model. We use a simulative approach similar to \cite{Fishbach:2019ckx} and make simplified assumptions to calculate the BBHs signal-to-noise ratio (\snr) and the uncertainties for the measure of the binary parameters. 
For each binary, we compute the detector chirp mass $\rm M_{chirp}$, symmetric mass ratio $\eta$, and luminosity distance $d_{\rm L}$.
The optimal \snr $\rho_{\rm true}$ of the binary is calculated with the approximation used in \citet{Fishbach:2019ckx,Mastrogiovanni_2021}:
\begin{equation}
        \rho_{\text{true}} = \rho_0 \, \theta \left( \frac{M_{\text{chirp}}}{M_{\text{chirp},0}} \right)^{5/6} \frac{d_{\rm L,0}}{d_{\rm L}},
        \label{eq: rho true}
\end{equation}
where $\rho_0=12$ is the reference \snr for a binary system with a chirp mass of $M_{\text{chirp},0}=26$ $\Msol$ and a luminosity distance $d_{\rm L,0}=1.9$ $\text{Gpc}$, consistent with the expected sensitivities for the O5 observing run \citep{Abbott_2020}. The projection factor $\theta$ accounts for the binary's orientation relative to the detector network, we defined it as the quadratic sum of the combined antenna response functions $F_+$ and $F_\times$ \citep{yunes}. 

To include the effects of noise across multiple detectors, we simulate an observed SNR, $\rho_{\rm det}$, by accounting for noise realizations in each detector. Specifically, we sample $\rho_{\rm det}$ using a non-central $\chi^2$ distribution with a non-centrality parameter $\rho_{\rm true}^2$ and $2 \times N_{\rm detectors} = 10$ degrees of freedom. In this simulation, we apply a detection threshold of $\rho_{\rm det}>12$. 

\begin{figure}
    \centering
    \includegraphics[width=0.5\textwidth]{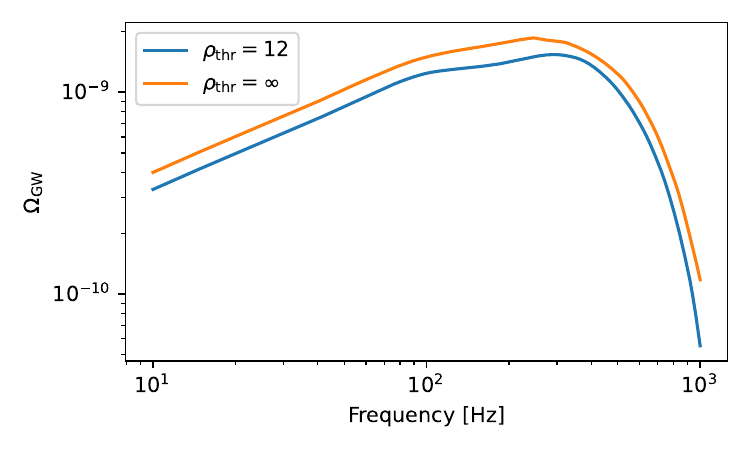}
    \caption{Stochastic gravitational-wave background calculated for the population of BBH sources (orange line) and after removing all the BBHs with \snr $> 12$ (blue line). The average SGWB power loss is about 30\% in all the sensitivity bands for ground-based GW detectors.}
    \label{fig:stoch}
\end{figure}

With the previous choices, we obtained that the resolvable sources correspond to about 0.6\% of the overall BBH population. In Fig.~\ref{fig:stoch}, we show the SGWB before and after removing the resolvable sources. As we can observe, for this simulative scenario, removing the resolvable individual sources would decrease the SGWB by $\sim 30\%$. As a consequence, the likelihood in Eq.~\ref{eq:combined likelihood} is not separable and that is the motivation for which our ``Combined SGWB and CBCs'' scenario splits two years of GW data into two independent datasets for the resolved sources and the SGWB sources.
The set ``Resolved CBCs'' contains about 1200 events detected, while the set ``Combined SGWB and CBCs'' contains about 600 solved events. 

Once detections are identified, posterior samples are generated using a likelihood model described as follows:
\begin{align}
    &\mathcal{L}({\rho_{\rm det}},  \log(M_{\text{chirp, det}}), \theta_{\rm det}, \eta_{det} | \rho, \log (M_{\text{chirp}}), \theta, \eta) = \nonumber\\
    &= \mathcal{L}(\rho_{\rm det} |\rho) \mathcal{L}(\log(M_{\text{chirp, det}})| \log(M_{\text{chirp}})) \mathcal{L}(\theta_{\rm det}|\theta) \mathcal{L}(\eta_{\rm det}|\eta),
    \label{eq: likelihood PE}
\end{align}
where $\mathcal{L}(\rho_{\rm det}| \rho)$ as we said follows a non-central $\chi^2$ distribution, while other for $M_{\text{chirp}}$, $\eta$, and $\theta$ we used the following normal distribution for the likelihoods:
\begin{align}
&\mathcal{L}(\log (M_{\text{chirp, det}})| \log(M_{\text{chirp}}))=\mathcal{N}\left(\log(M_{\rm chirp}), 0.08\frac{8}{\rho_{\rm det}}\right),\\
&\mathcal{L}(\theta_{\rm det}|\, \theta)= \mathcal{N}\left(\theta, 0.21\frac{8}{\rho_{\rm det}}\right) , \\
&\mathcal{L}(\eta_{\rm det}|\eta)= \mathcal{N}\left(\eta, 0.02 \frac{8}{\rho_{\rm det}}\right).
\end{align}

After obtaining posterior samples distributed according to this likelihood model with uniform priors, we converted them into source frame masses and luminosity distance using the SNR conversion Eq. \ref{eq: rho true}. Since there is an implied prior in this transformation, it is removed by calculating the Jacobian of the transformation from $(\rho_{\rm det}, M_{\rm chirp, det}, \eta_{\rm det}, \theta_{\rm det})$ to the final variables required for our analysis $(m_{\rm 1,s}, m_{\rm 2,s}, d_{\rm L})$.

\begin{figure}[h]
\centering
\includegraphics[width=0.4\textwidth]{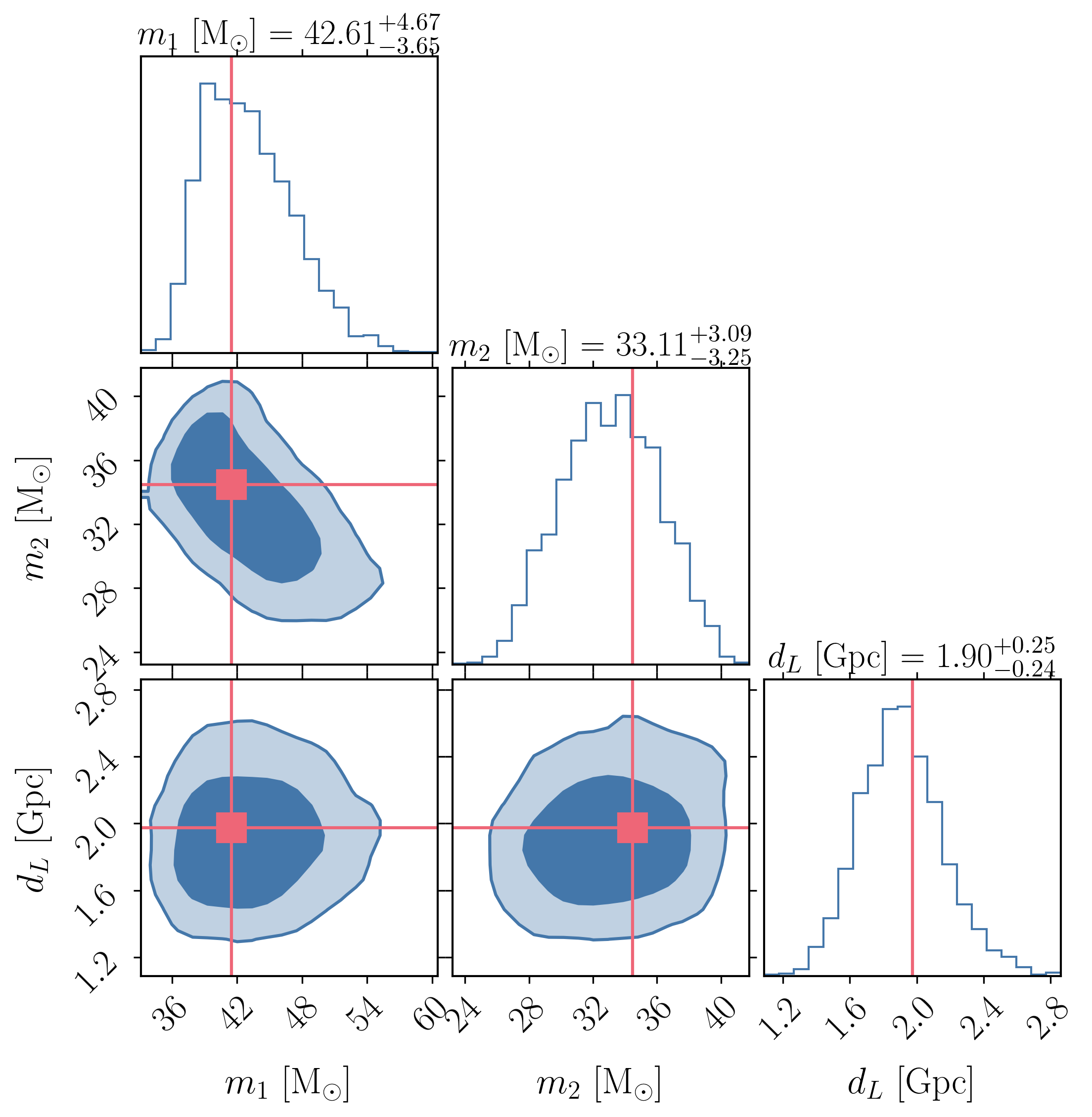}
\caption{Corner plot showing the posterior distributions of the key parameters estimated from the simulation (blue). The red lines represent the true values of the parameters injected into the simulation. The diagonal panels display the marginalized 1D posterior distributions for each parameter, while the off-diagonal panels show the 2D joint posterior distributions.}
\label{fig: corner_PE}
\end{figure}

An example of the posterior distributions for key parameters, including detector frame masses and luminosity distance, is shown in Fig. \ref{fig: corner_PE}. With this type of likelihood, we can see that it reproduces correlated mass estimates, as expected for real GW events. 

% Furthermore, in Fig. \ref{fig: scale relation dl rho det}, we show the luminosity distance uncertainty ratio $(\sigma_{d_L}/d_{\rm L}$) scales inversely with detected SNR, and we observe that for most events above the detection threshold SNR=12 the typical luminosity distance uncertainty is around 20\%, consistent with real observed events. However, as the detected SNR increases, it becomes possible to achieve more precise measurements. 
% \begin{figure}[h]
% \centering
% \includegraphics[width=0.4\textwidth]{Figures/SNR_vs_sigma_dL_2yr.png}
% \caption{Luminosity distance relative uncertainty $(\sigma_{d_L}/d_L$) as a function of detected SNR.}
% \label{fig: scale relation dl rho det}
% \end{figure}

\subsection{Stochastic Gravitational Wave Background}

We simulate cross-correlation measurements of the corresponding SGWB, assuming $T_{\rm obs} = 1$ yr of integration with LVK with O5 sensitivity and a coherence time for the stochastic search of 4 seconds. This corresponds to frequency point estimates of the cross-correlation statistic evaluated every 0.25 Hz. As a detector baseline we use a five-detector LIGO (Hanford, Livingston and India), Virgo and KAGRA networks at design sentivities\footnote{\url{https://dcc.ligo.org/LIGO-T2000012-v1/public}} \citep{Advanced_Virgo, Advanced_LIGO, Kagra, Ligo_India}.

\begin{figure}[h]
\centering
\includegraphics[width=0.45\textwidth]{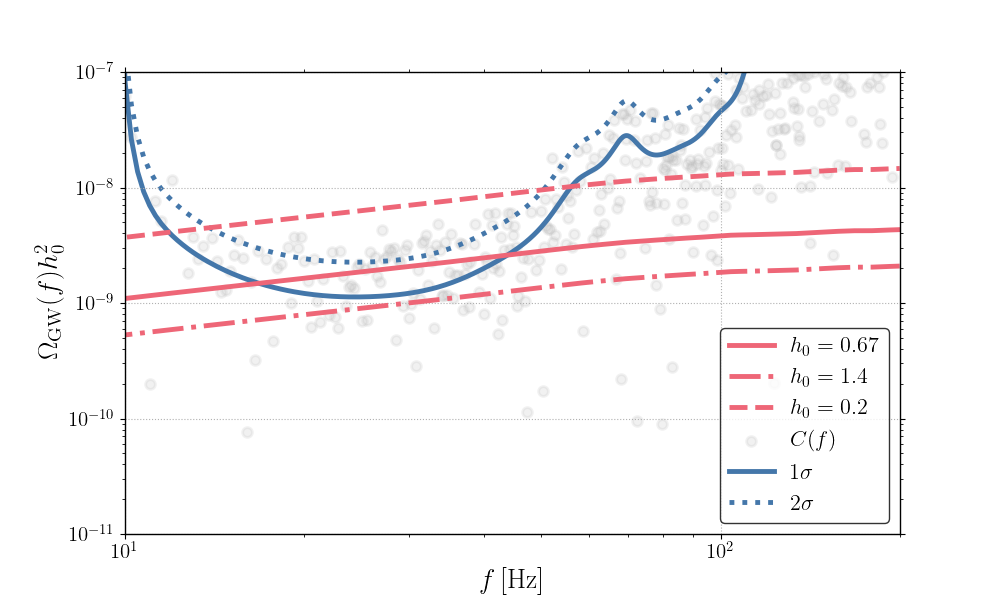}
\caption{Simulated SGWB cross-correlation spectrum $C(f)$ (points) with corresponding uncertainties $\sigma (f)$ (grey lines). The red linen are theoretical curves of $\Omega_{\rm GW}(f)$ for different $h_0^2$, and the blue curves display the $1\sigma$ and $2\sigma$ sensitivity of the LVK at the design sensitivity (1 year of observation in O5). The scatter reflects realistic fluctuations centered on the true $\Omega_{\rm GW}(f)$.}
\label{fig: sgwb simulated data}
\end{figure}

In Fig. \ref{fig: sgwb simulated data}, each data point is drawn from a Gaussian distribution centered on the true value of $\Omega_{\rm GW}(f)$, as calculated using Eq. \ref{eq: omega_gw}, with a standard deviation $\sigma(f)$ given by Eq. \ref{eq:combined variance}. The uncertainty $\sigma(f)$ represents the sensitivity to an SGWB in every frequency bin. As argued before, $\sigma(f)$ is directly related to the upper limits of the energy density of the SGWB, as fluctuation beyond this value is rare. This property allows us to detect the SGWB: if $\sigma(f)$ is much larger than $\Omega_{\rm GW}(f)$, the variations in $\Omega_{\rm GW}(f)$ become effectively indistinguishable within the noise, making detection unlikely. Conversely, when  $\sigma(f)$ is sufficiently small, deviations in $\Omega_{\rm GW}(f)$ can be resolved, enabling us to identify the SGWB signal.
Figure \ref{fig: sgwb simulated data} also shows why low $H_0$ values are easier to exclude with the SGWB. As we can observe, when we remove the $1/H_0^2$ dependency, namely when we normalize the critical density of the Universe $\rho_c$, the cross-correlation coefficients do not depend on $H_0$ (as also indicated in Eq.~\ref{eq:cross-corr estimator general}), while the SGWB does. As a result, the SNR for the SGWB increases as $1/H_0$.

\section{Results with simulated data}
\label{sec:Result}

Using the simulated data described in Sec.~\ref{sec:Data} and prior ranges adopted are listed in Tab. \ref{tab:priors}. We perform three separate analyses. The first considers only resolved CBC events (``Resolved CBCs''), using two years of data. The second is a combined analysis (``Combined SGWB and CBCs''), where we use one year of resolved BBH detections and one year of SGWB measurements. Finally, the third analysis focuses solely on one year of the SGWB to assess its sensitivity and, if possible, directly observe the impact of improvements due to unresolved sources.

\begin{figure}[h]
\centering
\includegraphics[width=0.5\textwidth]{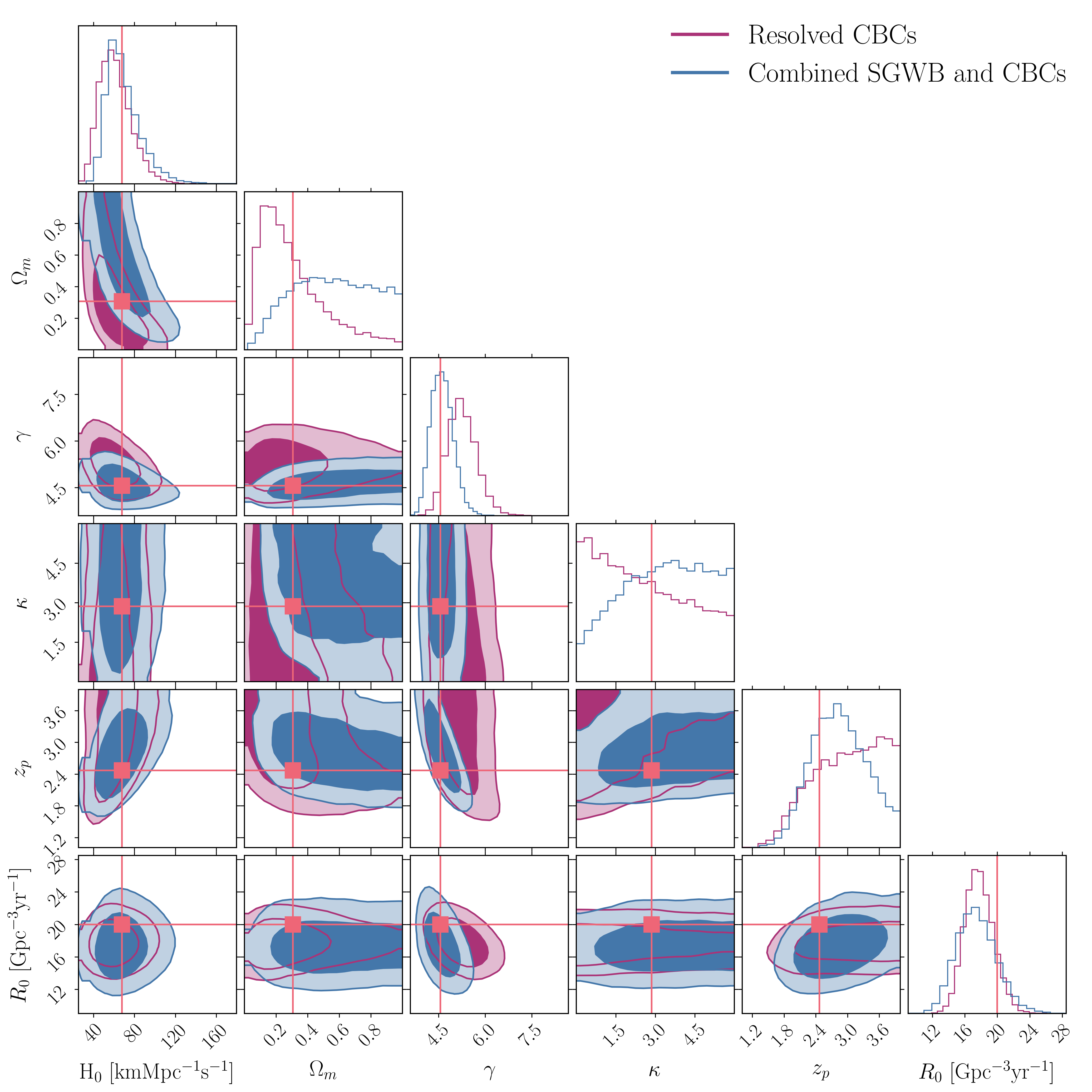}
\caption{Posterior on the population parameters governing the cosmic expansion and the CBC merger rate as a function of redshift for the data set ``Resolved CBCs'' (2yrs) of observations and the data set ``Combined SGWB and CBCs'' (1 yr for individual sources and 1 yr for the SGWB). The injected values are marked with the orange dot.}
\label{fig: corner rate}
\end{figure}

Figure \ref{fig: corner rate} displays the corner plot for the cosmological parameters $H_0$ and $\Omega_{\rm m}$ with the redshift rate parameters. From the analysis using only resolved CBCs, we find $H_0 = 59.7^{+17.1}_{-13.4} \ \text{km s}^{-1} \text{Mpc}^{-1}$ at 68\% credible interval (CI). Combining SGWB information we find $H_0 = 66.2^{+19.3}_{-12.1} \ \text{km s}^{-1} \text{Mpc}^{-1}$ at 68\% CI, indicating no significant improvement in precision. This result indicates that most of the precision on $H_0$ is determined by resolved spectral sires. However, we note that the inclusion of the SGWB significantly helps in excluding the region of the parameter space that covers low values of $H_0$ and $\Omega_m$. As also indicated by \citet{cousins_2025}, this is a consequence of the fact that the SNR for the SGWB detection is higher for low values of $H_0$ and $\Omega_m$.

%An interesting finding is that the inclusion of the SGWB helps in breaking the degeneracy between $H_0$ and the matter-energy fraction $\Omega_m$ and $R_0$. For resolved sources, the $H_0-R_0$ correlation is introduced by the fact that $\gamma$ is correlated with $H_0$ and $R_0$ are anti-correlated with $\gamma$ as more GW signals could be either explained by increasing $R_0$ or increasing $\gamma$.  While the correlation with $\Omega_m$ arises from the fact that these two cosmological parameters govern together the conversion between luminosity distance and redshift. 

We observe that both the data sets for individual sources and SGWB measure the rate parameter $R_0$ with similar precision. This is a consequence of the fact that the individual sources are mostly detected at lower redshifts and are the ones driving the inference on these population parameters. On $\gamma$ we achieve an additional $10\%$ precision with the combined analysis. 
We also see that the inclusion of the SGWB slightly introduces some preferences for the posteriors on $z_p$ and $k$. These two parameters govern the peak of the BBH merger rate and its redshift evolution for $z>z_p$. The inclusion of the stochastic background makes possible the measurement $z_p = 2.8^{+0.6}_{-0.5}$ and helps to exclude low values of $k$, namely, we exclude models in which the rate increases beyond $z_p$. 

\begin{figure*}[h]
\centering
\includegraphics[width=\textwidth]{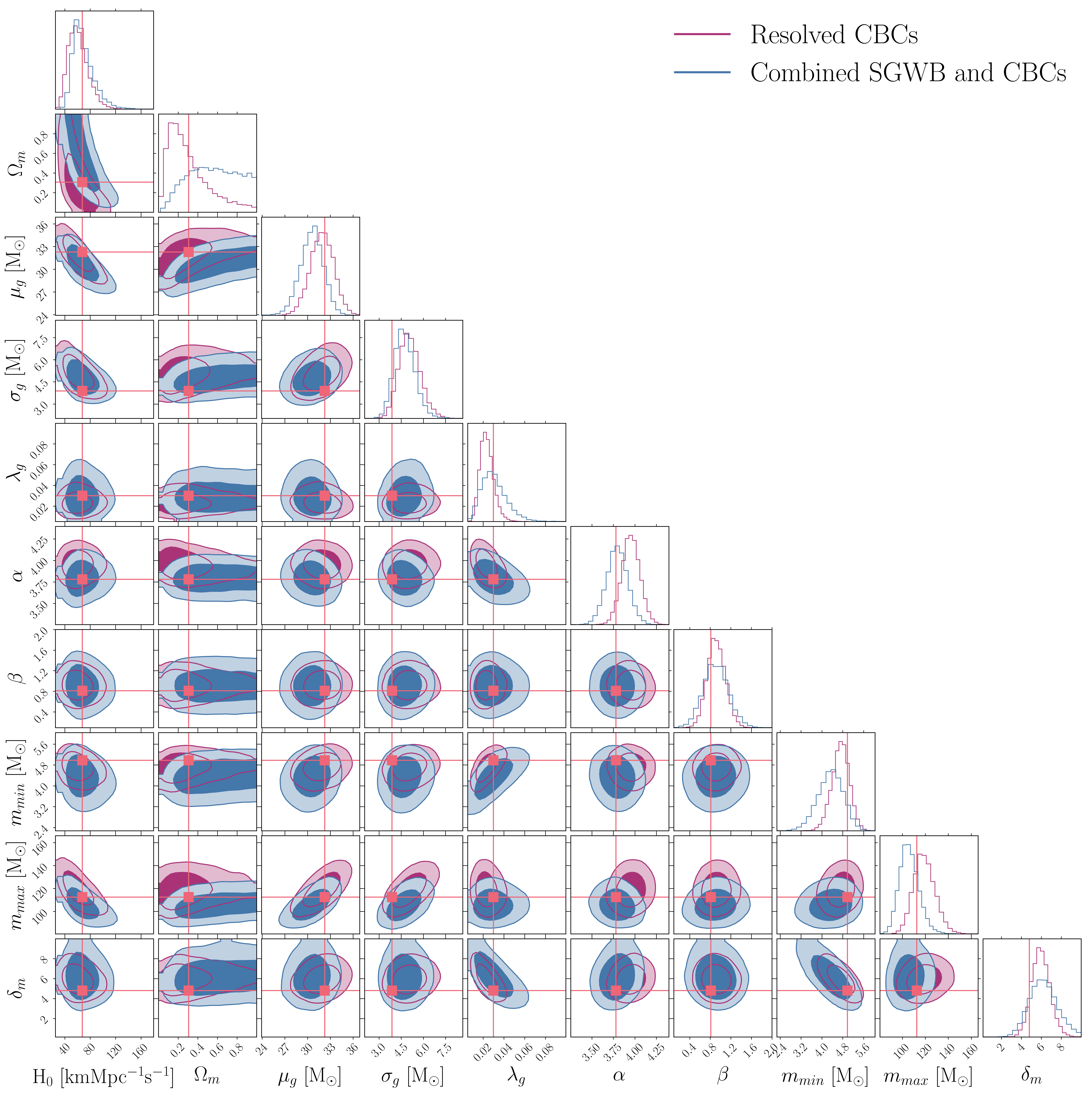}
\caption{Posterior on the population parameters governing the cosmic expansion and the CBC mass parameters for the data set ``Resolved CBCs'' (2 yrs) of observations and the data set ``Combined SGWB and CBCs'' (1 yr for individual sources and 1 yr for the SGWB). The injected values are marked with the orange dot.}
\label{fig: corner mass model}
\end{figure*}

Figure \ref{fig: corner mass model} focuses on the mass model parameters and correlations with $H_0$. The combined analysis does not lead to any improvement in these parameters, whereas the resolved-only analysis performs better.

\begin{figure*}
    \centering
    \includegraphics[width=\textwidth]{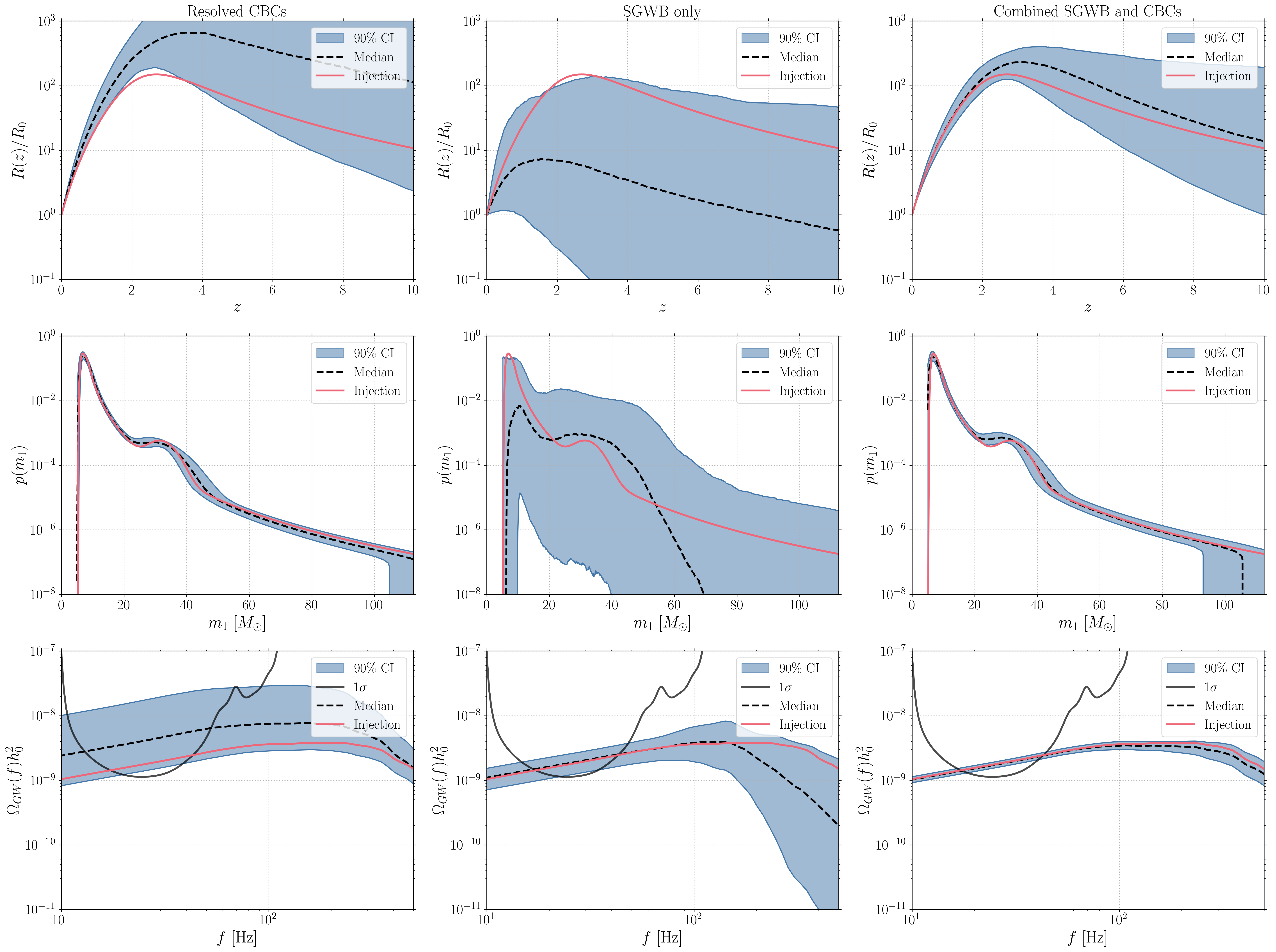}
    \caption{Posterior predictive checks are shown for resolved CBC (left), SGWB-only analysis (center), and the joint analysis of both sources (right). The dashed and solid black curves show the median and 90\% credible bounds and the red curve is the fiducial value injected. The top panel shows the rate density $ R(z)$ of binary black hole mergers over the local merger rate $R_0$ as a function of redshift. The middle panel shows the primary mass distribution as a function of the primary mass. The bottom panel shows the posterior on $\Omega_{GW}h_0^2$ as a function of the frequency. In these panels, the black line represents the 1$\sigma$ curve. }
    \label{fig: posterior predictive checks}
\end{figure*}

To assess the performance of the inference, we produce posterior predictive checks (PPCs), which are defined as:
\begin{align}
&p(\theta|{x}) = \int p(\Phi|{x}) p(\theta|\Phi) d\Phi, \\
&\Omega_{\text{GW}}(f|{x}) = \int p(\Phi|{x}) \Omega_{\text{GW}}(f;\Phi) d\Phi,
\end{align}
where $p(\Phi|{x})$ represents the posteriors on the population parameters and $\{x\}$ the GW data that can either include or not the SGWB.  
By analyzing the PPCs for all three analyses in Fig. \ref{fig: posterior predictive checks} we gain insight into the reconstructed redshift distribution (top panel), mass distribution (middle panel), and SGWB energy density (bottom panel), $\Omega_{\text{GW}}(f)$. 
In Fig.~\ref{fig: posterior predictive checks} we report the posterior predictive checks for the analyses that we have performed. 
When considering resolved sources alone, the reconstructed redshift distribution is precise at low redshifts, while at higher redshifts $z \geq z_p$, the PPC is highly prior-dominated as we can not determine the values of $k$ and $z_p$. When using the SGWB-only data, the reconstruction of the rate is strongly prior-dominated as there is too much degeneracy among the rate parameters $R_0, \gamma$ and $k$, although there appears to be some information on the values of $z_p$. Including individual sources and SGWB together, provides a reconstruction of the rate informative up to redshift $z=2-3$ consistent with what is observed in \citep{Callister_2020}. As we can see, by combining the resolved sources and the SGWB, the reconstruction of the BBH merger rate is less prior-dominated.

For the mass spectrum, we obtain that most of the information is collected from individual sources. In fact, for the SGWB-only analysis, the reconstruction of the mass spectrum is strongly prior dominated. The improvement in the precision of the reconstruction of SGWB energy density, when considering individual and SGWB sources, is mostly due to the improvement in the measure of $\gamma$ and $z_p$ implied by the CBC rate parameters.

% \begin{figure}[h]
% \centering
% \includegraphics[width=0.4\textwidth]{Figures/corner_H0_Om0_2yr.png}
% \caption{Expected posteriors on all cosmological $ H_0$ and $\Omega_m$ using LVK observation at O5 sensitivity. Purple distributions (both one- and two-dimensional) are the results obtained using a mock catalog of direct CBC detections; blue ones correspond to the synthesis of the CBC catalog with simulated stochastic measurements. The red line is the fiducial value injected in the simulation.}
% \label{fig:corner cosmo}
% \end{figure}

\begin{table*}[h!]
    \centering
    \caption{Overview of population hyperparameters and prior choices. $U(\cdot)$ represents a uniform distribution. The \textsc{Power law + peak} model is used only for the simulations in Sec.~\ref{sec:Result}, while the \textsc{Multi + peak} model is used for the analysis of O3 data in Sec.~\ref{sec:Result_data}. The priors used on cosmological parameters and rate evolution parameters are the same for the simulation and real data cases. For the simulation case, we also report the injected value. The rate parameters, mass model parameters and $H_0$ are chosen to be consistent with the posterior distributions from \citep{GWTC3_cosmo} while $\Omega_m$ is set by \citep{2016}.}
    \begin{tabular}{@{}llll@{}}
        \toprule
        \textbf{Parameter} & \textbf{Definition} & \textbf{Injected Value}& \textbf{Prior Range} \\
        \midrule
        \multicolumn{3}{l}{\textbf{Cosmological Parameters (flat $\Lambda$CDM)}} \\
        $H_0$ & Hubble parameter [km s$^{-1}$ Mpc$^{-1}$] & 67.7& $U(10, 200)$ \\
        $\Omega_{m}$ & Matter density parameter & 0.3065 & $U(0, 10)$ \\
        \midrule
        \multicolumn{3}{l}{\textbf{Rate Evolution (Madau-rate)}} \\
        $\gamma$ & Evolution slope for $z < z_p$ & 4.56 & $U(0, 12)$ \\
        $\kappa$ & Evolution slope for $z > z_p$ & 2.86 &$U(0, 6)$ \\
        $z_p$ & Redshift of peak rate & 2.47 & $U(0, 4)$
        \\
        $R_0$ & Local merger rate [$\text{Gpc}^{-3}\text{yr}^{-1}$] &20 &$U(0, 100)$ \\
        \midrule
        \multicolumn{3}{l}{\textbf{Mass Function Parameters (PowerLaw+Peak)}} \\
        $\alpha$ & Power-law index of primary mass & 3.78& $U(1.5, 12)$ \\
        $\beta$ & Power-law index of secondary mass & 0.81 & $U(-4, 12)$ \\
        $\delta_m$ & Smoothing parameter [$M_\odot$] & 4.8& $U(0, 10)$ \\
        $m_{\text{min}}$ & Minimum source mass [$M_\odot$] & 4.98 & $U(2, 10)$ \\
        $m_{\text{max}}$ & Maximum source mass [$M_\odot$] & 112.5& $U(50, 200)$ \\
        $\mu_g$ & Mean of the Gaussian peak [$M_\odot$] & 32.27& $U(20, 50)$ \\
        $\sigma_g$ & Width of the Gaussian peak [$M_\odot$] & 3.88 & $U(0.4, 10)$ \\
        $\lambda_g$ & Fraction of events in  the Gaussian peak & 0.03 & $U(0, 1)$ \\
        \midrule
        \multicolumn{3}{l}{\textbf{Mass Function Parameters (Multi + peak)}} \\
        $\alpha$ & Power-law index of primary mass & & $U(1.5, 12)$ \\
        $\beta$ & Power-law index of secondary mass & &$U(-4, 12)$ \\
        $m_{\text{min}}$ & Minimum source mass [$M_\odot$] & & $U(2, 10)$ \\
        $m_{\text{max}}$ & Maximum source mass [$M_\odot$] & & $U(50, 200)$ \\
        $\delta_m$ & Smoothing parameter [$M_\odot$] & & $U(0, 10)$ \\
        $\mu_{g,\text{low}}$ & Mean of the lower Gaussian peak [$M_\odot$] &  & $U(7, 12)$ \\
        $\mu_{g,\text{high}}$ & Mean of the higher Gaussian peak [$M_\odot$] &  & $U(20, 50)$ \\
        $\sigma_{g,\text{low}}$ & Width of the lower Gaussian peak [$M_\odot$] & & $U(0.4, 5)$ \\
        $\sigma_{g,\text{high}}$ & Width of the higher Gaussian peak [$M_\odot$] & & $U(0.4, 10)$ \\
        $\lambda_g$ & Fraction of events in the peaks & & $U(0, 1)$ \\
        $\lambda_{g,\text{low}}$ & Fraction of events in the lower peak & & $U(0, 1)$ \\
        \bottomrule
    \end{tabular}
    \label{tab:priors}
\end{table*}

\section{Reanalysis of GWTC-3}
\label{sec:Result_data}

Using the approach defined in Sec.~\ref{sec:Analysis Method} and the likelihood in Eq.~\ref{eq:combined likelihood}, we reperform a population and cosmological spectral siren run with data from the latest public LVK observing run. 

We select 59 BBHs detected during O3 with an inverse false alarm rate higher than 1 yr and we use their posteriors released with GWTC-3 \citep{GWTC_3}. To correct selection biases for the Poisson part of the likelihood, we use a set of injections in real O3 data\footnote{\url{https://zenodo.org/records/5636816}} released with \citet{PhysRevX.13.011048}.
For the stochastic data, we use the frequency point estimation of $\hat{C}(f)$ and $\sigma^2(f)$ released\footnote{\url{https://dcc.ligo.org/LIGO-G2001287/public}} with \citet{PhysRevD.104.022004}. We only consider frequencies below $200 \ \rm Hz$ to ease the computational load of the analysis. The $\hat{C}(f)$ that we use are evaluated every $0.03 \ \rm Hz$, corresponding to a coherence time for the SGWB search of $32 \ s$ and they are calculated with $H_0 = 67.9 \ {\rm km s^{-1} Mpc^{-1}}$. 
For this analysis, we still use the same BBH merger redshift model considered for the simulations. While for the mass model, we adopt a \textsc{Multi peak} model, that can describe also a possible feature at masses $\sim 10 M_\odot$. The \textsc{Multi peak} model describes the mass distributions as 
\begin{align}
    & p(m_{\rm 1,s} | m_{\text{min}}, m_{\text{max}}, \alpha) = (1 - \lambda_\text{g}) P(m_{\rm 1,s} | m_{\text{min}}, m_{\text{max}},\alpha) +\\
    & + \lambda_\text{g} \lambda_\text{g,low} G(m_{\rm 1,s} | \mu_{\rm g,low}, \sigma_{\rm g,low}) \nonumber  \\
    & + \lambda_\text{g} (1-\lambda_\text{g,low}) G(m_{\rm 1,s} | \mu_{\rm g,high}, \sigma_{\rm g,high}), \nonumber  \\
    &p(m_{\rm 2,s} | m_{\text{min}}, m_{\rm 1,s}, \beta) = P(m_{\rm 2,s} | m_{\text{min}}, m_{\rm 1,s}, \beta).
    \label{eq:mass model MLP}
\end{align}
For the reanalysis of O3, we also make the implicit assumption that the likelihood in Eq.~\ref{eq:combined likelihood} is separable. For the O3 runs, this is a reasonable assumption as the removal of the individual sources of the population does not modify significantly the SGWB \citep{PhysRevD.104.022004}. The priors used for the analysis are reported in Tab.~\ref{tab:priors}.

We perform two analyses, one considering only individual sources and the other considering also the SGWB upper limits. We find that the constraints on the population parameters, including also the cosmological parameters, are entirely given by the individual GW sources.
\begin{figure}
    \centering
    \includegraphics[width=1.\linewidth]{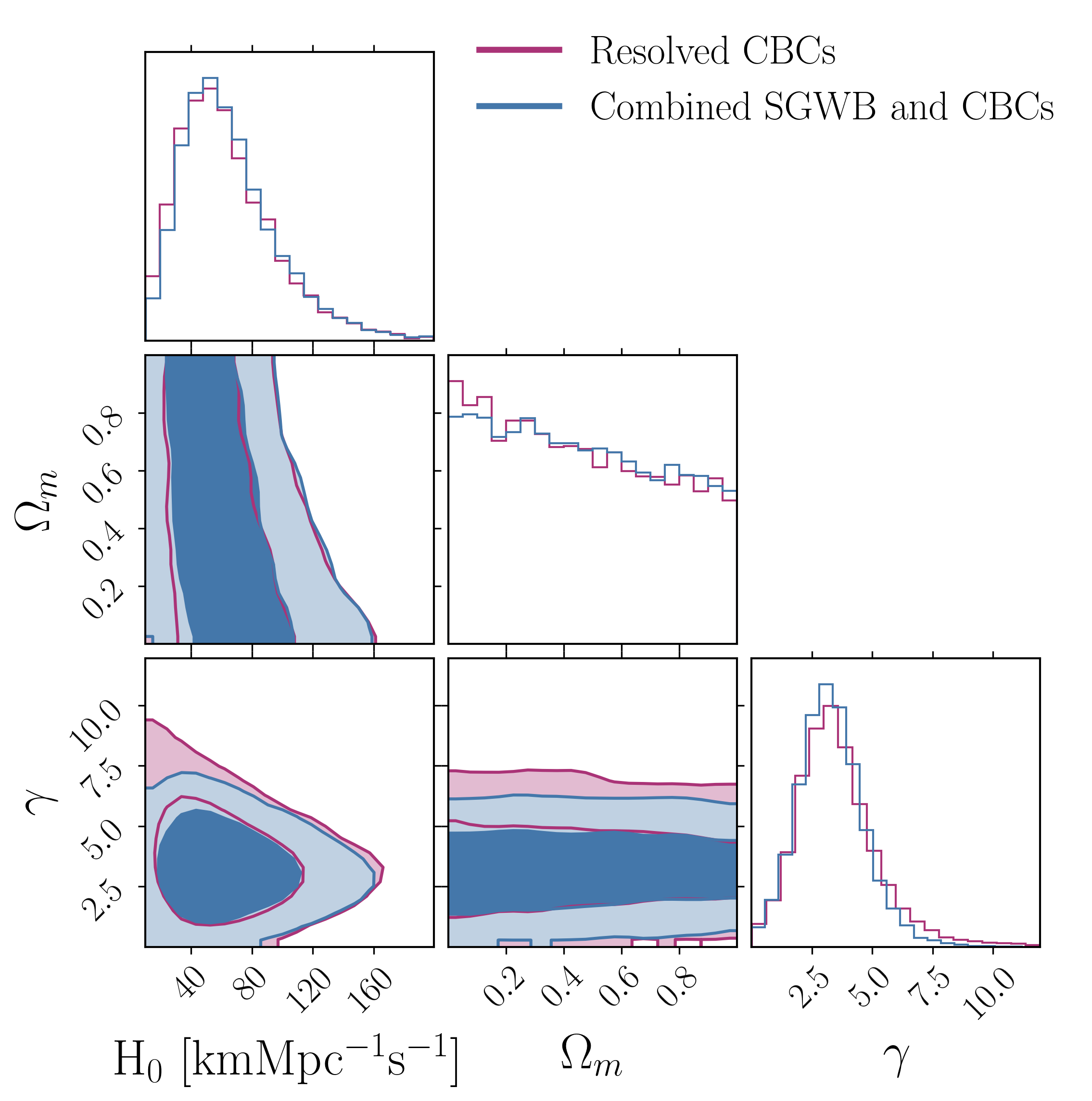}
    \caption{Marginalized joint posterior distribution on the cosmological expansion parameters, and the rate parameter $\gamma$ for an analysis including only 59 BBHs detected in O3 (purple line) and considering also the SGWB (blue line). The contour marks the $1\sigma$ and $2\sigma$ CIs.}
    \label{fig:GWTC3corner}
\end{figure}
In Fig.~\ref{fig:GWTC3corner}, we display the marginalized joint posterior distributions for these two analyses on the Hubble constant, $\Omega_m$ and the parameter $\gamma$ for the rate evolution. We find that the inclusion of the SGWB weakly helps in excluding low values of $H_0$ and high values of $\gamma$ from the  $2\sigma$ CI areas. The posterior on the remaining of the population parameters is unchanged and we do not display it.
\begin{figure*}
    \centering
    \includegraphics[width=0.8\linewidth]{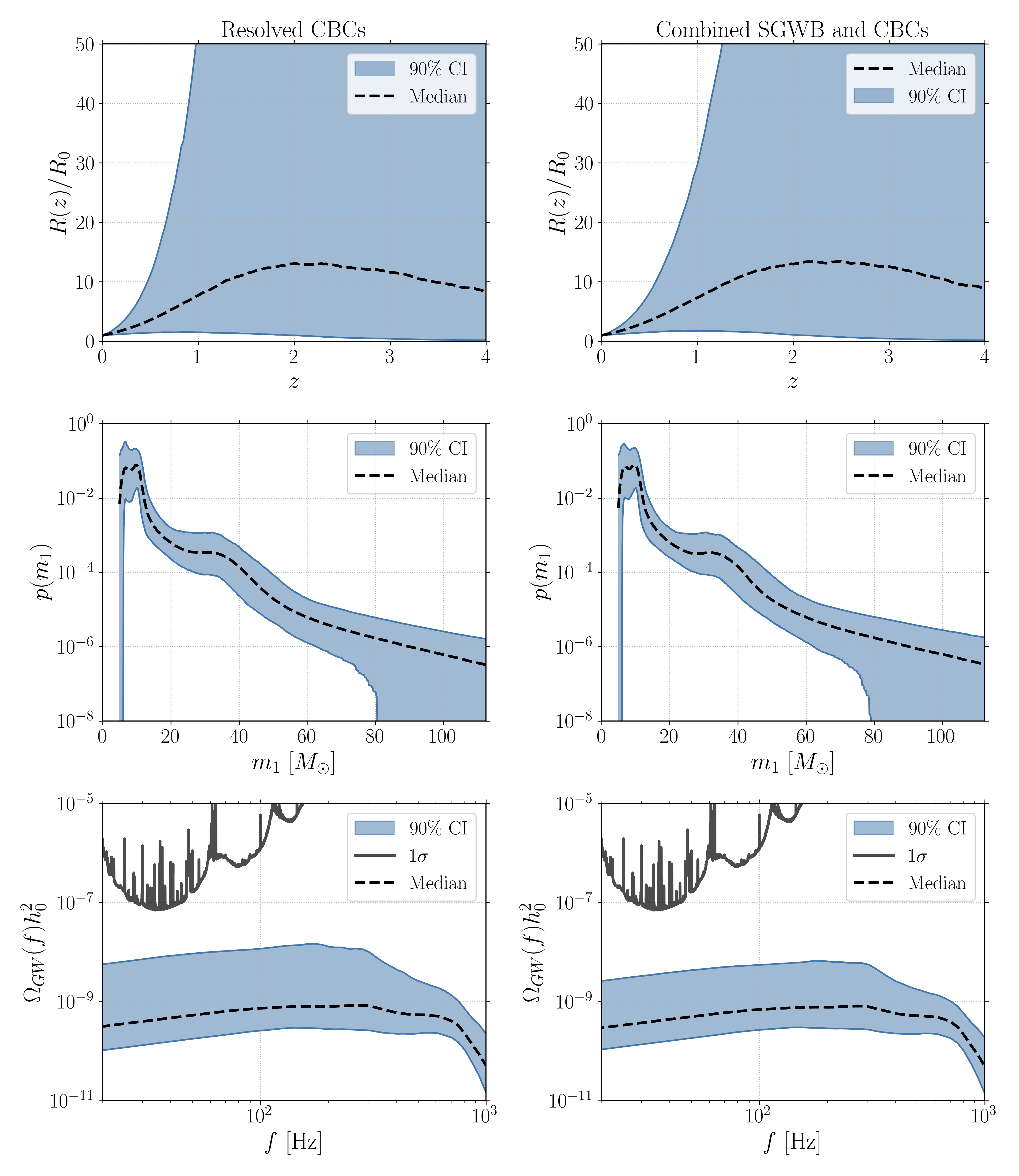}
    \caption{Posterior predictive checks are shown for resolved CBC (left), and the joint analysis of both sources (right). The dashed and solid black curves show the median and 90\% credible bounds and the red curve is the fiducial value injected. The top panel shows the rate density $ R(z)$ of binary black hole mergers over the local merger rate $R_0$ as a function of redshift. The middle panel shows the primary mass distribution as a function of the primary mass. The bottom panel shows the posterior on $\Omega_{GW}h_0^2$ as a function of the frequency. In these panels, the black line represents the 1$\sigma$ curve. }
    \label{fig:posterior_predictive_GWTC3}
\end{figure*}
In Fig.~\ref{fig:posterior_predictive_GWTC3}, we show the PPCs for the BBHs rate, mass distribution and SGWB reconstructed from the two analyses. We notice that the mass and rate reconstructions are equivalent in the case that the SGWB is included or not. We can also see that the reconstructed SGWB in both analyses is well below the $1\sigma$ upper limits from the stochastic searches. This means that the SGWB prediction is entirely given by the individual sources of CBCs and no additional information can be included from the SGWB likelihood. This is consistent with what was found in \citet{Callister_2020}.

In parallel to this work, \citet{cousins_2025} presented a spectral siren analysis using GWTC-3 including also the SGWB. Although the results that we obtained are in excellent agreement, namely the inclusion of the SGWB does not modify significantly the inference on population and cosmological parameters, there are few technical differences in our approach. 
From a data selection point of view, \citet{cousins_2025} uses 42 BBH events with \snr$>11$ observed in the first three LIGO and Virgo observing while we consider 59 BBHs detected during O3 with inverse false alarm rate higher than 1 per year. Another difference is that \citet{cousins_2025} uses a \textbf{Power Law + peak} model to describe the masses while we use a \textbf{Multi Peak} model.
From a data analysis perspective, \citet{cousins_2025} uses as a prior on the population and cosmological parameters the posterior in \citet{GWTC3_cosmo} from the spectral siren analysis. This is motivated indeed by the separability of the likelihood in Eq.~\ref{eq:combined likelihood} that we also assume in our framework. Then the SGWB information is included as a likelihood term fitting posterior samples from \citet{PhysRevD.104.022004} on $\Omega_{\alpha=2/3}$, which is a quantity describing a SGWB as 
\begin{equation}
\Omega_{\rm GW} (f) = \Omega_{\alpha=2/3} \left(\frac{f}{f_{\rm ref}}\right)^{2/3},
\label{eq:presc}
\end{equation}
and corresponds to the SGWB expected by a population of BBHs emitting GWs during their inspiral phase at the 0 Post-Newtonian order. Instead, we calculate $\Omega_{\rm GW}$ reweighting a GW energy spectrum that is calculated considering corrections for the GW emission up to the 3.5 Post-Newtonian order \citep{Ajith:2007kx} and also for the merger and ring-down phases. However, we notice that our two approaches reach equivalent results as the SGWB is dominated in the $20-200 \ \rm Hz$ region by the inspiral of BBH merger that can be safely estimated with the 0 Post-Newtonian order emission, therefore in that region, the SGWB is well approximated by the prescription in Eq.~\ref{eq:presc}.

\section{Conclusions}
\label{sec:conc}

In this work, we discussed how it is possible to include the SGWB into spectral siren analyses for GW cosmology and enhance our understanding of astrophysical populations and cosmological parameters. Using simulated data at design sensitivity of the LVK network’s O5 observing run, we applied a hierarchical Bayesian framework to infer a set of 14 hyperparameters $\Phi$, constraining key cosmological parameters such as $H_0$ and $\Omega_m$, along with parameters governing the CBC mass distribution and the redshift evolution of the merger rate.

Our analysis shows that combining SGWB data into the inference framework does not improve the precision on the marginal posteriors for cosmological parameters. 
In particular, we see that the Hubble constant estimates derived solely from resolved events yielded $H_0 = 59.7^{+17.1}_{-13.4} \ \text{km s}^{-1} \text{Mpc}^{-1}$ at 68\% CI, while combining the SGWB information yielded $H_0 = 66.2^{+19.3}_{-12.1}  \ \text{km s}^{-1} \text{Mpc}^{-1}$ at 68\% CI, indicating no significant improvement in precision. However, we obtain that the inclusion of the SGWB helps in excluding the lowest values of $H_0$ and $\Omega_m$. 
Besides cosmology, on an astrophysical population-level, the integration of SGWB data slightly improves our understanding of the redshift evolution of CBC rates, it helps the measurements of $\gamma$ and $z_p$. The SGWB's sensitivity to unresolved sources at higher redshifts allows for tighter constraints at high redshifts of the merger rate distribution.

When analyzing real O3 data, we find that the inclusion of the SGWB does not significantly improve any of the population or cosmological parameters. This result is consistent with the same findings of \citet{Callister_2020, Abbott_2021} for population studies and \citet{, cousins_2025} for the cosmological expansion parameters, and it is a consequence of the fact that the SGWB implied by individual sources is well below the sensitivity estimates for current stochastic searches.

Of course, in this exploratory study, we made a few simplistic assumptions that can be revised for follow-up studies. First, to preserve the separability of the likelihood in Eq.~\ref{eq:combined likelihood}, we made the suboptimal choice of splitting the data disjoint sets used either for the SGWB or individual sources searches. This assumption can be improved by developing a likelihood that takes into account additional correlations on the detected sources.  In this sense, we expect our results for the improvement on the determination of the cosmological parameters to be ``conservative'', as we are not using the full information present in the data set. We expect that, with a joint likelihood for resolved sources and SGWB, the precision on the cosmological parameters could be improved when including the stochastic.
We also did not consider the possible contribution for the SGWB introduced by binary neutron stars and neutron star black holes binaries, that can be modeled with multi-population models.

\begin{acknowledgements}
We are grateful to M.~Mancarella, A.~Renzini and A. Toubiana for discussions during the development of this work.
We are grateful to \citet{cousins_2025} and collaborators for the constructive feedback received during the internal LVK review of this work.
This work received support from the French government under the France 2030 investment plan, as part of the Excellence Initiative of Aix Marseille University - amidex (AMX-19-IET-008 - IPhU).
SM is supported by ERC Starting Grant No. 101163912–GravitySirens.
This research made use of data, software and/or web tools obtained from the Gravitational Wave Open Science Center, a service of the LIGO Scientific Collaboration, the KAGRA Collaboration and the Virgo Collaboration.
This material is based upon work supported by NSF's LIGO Laboratory which is a major facility fully funded by the National Science Foundation.
The authors are grateful for computational resources provided by the LIGO Laboratory (LHO) and supported by National Science Foundation Grants PHY-0757058 and PHY-0823459.

\end{acknowledgements}

\begin{appendix}
\section{Computation of Hierarchical and Stochastic Likelihoods}
\label{appendix A}
We use the Python package \icarogw to analyze our data. For a detailed explanation of its functionality, refer to \citet{mastrogiovanni2023icarogwpythonpackageinference}. Here, we provide a summary of the likelihood computation. The main application of \icarogw is to infer the population parameters $\Phi$ that describe the production rate of events in terms of the GW parameters $\theta$, namely $ \frac{\de N}{\de t\de\theta} (\Phi)$. The hierarchical likelihood in Eq. \ref{eq: hierarchical likelihood}, introduced in Section \ref{subsec:Hierarchical Bayesian Likelihood}, models the probability of observing $ N_{\rm obs}$ observations, each described by some parameters $\theta$, in a dataset $\{x\}$ over an observing time $ T_{\rm obs}$ accounting for selection biases \citep{Vitale_2021}. \icarogw computes numerically the hierarchical likelihood in Eq. \ref{eq: hierarchical likelihood} as:
\begin{equation}
    \ln [\mathcal{L}(\{x\}|\Phi)]\approx -\frac{ T_{\rm obs}}{ N_{\rm gen}}\sum^{N_{\rm det}}_{j=1}s_j +\sum^{ N_{\rm obs}}_{i=1} \ln \left[\frac{T_{\rm obs}}{N_{ s,i}}\sum^{N_{s,i}}_{j=1}w_{i,j}\right],
\end{equation}
where $w_{i, j}$ and $s_j$ are weights: $w_{ i, j}$ has a dimension equal to the number of CBC mergers happening per unit of time and weights the sample, while $s_j$ weights the injections and it is defined with the dimension of a CBC merger rate per detector time.  The first term derives from Monte Carlo integration to estimate:
\begin{equation}
    N_{\rm exp}(\Phi)=T_{\rm obs}\int  p_{\rm det}(\theta, \Phi) \frac{\de N}{\de t \de\theta}\de\theta,
\end{equation}
where $p_{det}(\theta, \Phi)$ is a detection probability or selection bias. Since an analytical form of $p_{\rm det}(\theta, \Phi)$ is typically unavailable, selection biases are evaluated using Monte Carlo simulations of injected and detected events \citep{Abbott_2021}. \icarogw takes in input a set of $N_{\rm det}$ detected injections out of $N_{\rm gen}$ total injections generated from a prior $\pi_{\rm inj} (\theta)$ to calculate the selection bias.

For each population model, \icarogw calculates two numerical stability estimators. The first is the effective number of posterior samples per event:
\begin{equation}
    N_{{\rm eff},i}=\frac{\left(\sum^{ N_{s,i}}_{j=1} w_{i,j}\right)^2}{\sum^{N_{s,j}}_{j=1}w^2_{i,j}},
    \label{eq: expected number of events}
\end{equation}
which ensures sufficient samples contribute to the integral, with a typical threshold of at least 20 effective samples per event. The second is the effective number of injections:
\begin{equation}
    N_{\rm eff, inj}=\frac{\left[\sum^{N_{\rm det} }_{j=1}s_j\right]^2}{\left[\sum^{N_{\rm det} }_{j=1}s^2_j -N_{\rm gen}^{-1}\left(\left[\sum^{N_{\rm det} }_{j=1}s_j\right]\right)^2\right]},
    \label{eq: effective number of injections}
\end{equation}
where numerical stability typically requires $ N_{\rm eff,inj} > 4N_{\rm obs}$.

The likelihood computation involves technical flags for configuration: \texttt{nparallel}=2048, which is the number of posterior samples that will be used per event to compute $w_{i, j}$, $\texttt{neffPE}=-1$, which is the effective number of posterior samples {PE} per event in Eq. \ref{eq: expected number of events} and \texttt{neffINJ=None}, which is the effective number of injections required by the hierarchical likelihood. If \texttt{neffINJ=None}, a default threshold of \texttt{NeffINJ}=$4\times N_{\rm obs}$ will be used.

We developed and implemented in \icarogw a new function to calculate the SGWB likelihood, as described in Eq. \ref{eq: stochastic likelihood}, optimizing the computation of $\Omega_{\rm GW}$. The expected $\Omega_{\text{GW}}(f)$ calculation, conditioned on a set of hyperparameters $\Phi$, follows the form outlined in Eq. \ref{eq: omega_gw}. This value depends on the average energy $\left\langle \frac{ \de E}{{\de}f_s} \right\rangle$ radiated by individual binaries, which is expressed as:
\begin{equation}
    \left \langle \frac{\de E}{{\de}f_s} \right\rangle=\int \de\Phi p(\Phi)\frac{\de E_s}{{\de}f_s}(\Phi),
\end{equation}
where $\Phi$ denotes the intrinsic properties of a given binary (masses, redshift, etc.), $p(\Phi)$ is the distribution of these parameters across the CBC population and this quantity is evaluated at the source-frame frequency $f_s = f (1 + z)$ in Eq. \ref{eq: omega_gw}. The GW energy spectrum is calculated with functionalities from \textsc{pygwb} \citet{Renzini:2023qtj} and \citet{Turbang_2024} that calculated the GW energy spectrum up to the 3.5 Post-Newtonian order for the GW emission and the merger and ring-down phase for non-precessing binaries.

Here, the integration spans the component masses $m_1$ and $ m_2$. Extending this to include integration over redshift, the computation of the expected $\Omega_{\text{GW}}(f)$ becomes a three-dimensional integral. A practical alternative is to use Monte Carlo integration following the methodology used in \cite{Turbang_2024}, which estimates the integral by averaging over numerous samples drawn from the population.

In the implementation of this approach, we start by generating $N$ random samples from uniform distributions for the parameters $ z$, $m_1$, and $ m_2$. Let the $i$-th sampled set of parameters be denoted as $\{z_i, m_{1,i}, m_{2,i}\}$. For each of these samples, we calculate the energy spectrum denoted as $\left(\frac{\de E}{{\de}f_s}\right)_i$. However, to evaluate $\Omega_{\text{GW}}(f)$ for the population described by the hyperparameters $\Phi$, reweighting of these samples is necessary. The computation is expressed schematically as:
\begin{equation}
\Omega_{\text{GW}}(f) \propto \frac{1}{\rm N} \sum_{i=1}^{N} { w_i} \left(\frac{\de E}{{\de}f_s}\right)_i,
\end{equation}
where $w_i$ are weights defined as:
\begin{equation}
     w_i \propto \frac{\mathcal{R}(z_i) (1+z_i) H(z_i)^{-1} p(m_{1i}) p(m_{2i})}{p_{\text{draw}}(z_i) p_{\text{draw}}(m_{1i}) p_{\text{draw}}(m_{2i})}.
\end{equation}
Here, $p_{\text{draw}}$ refers to the uniform distributions from which the initial samples were drawn, and the numerators represent the target distributions associated with the population characterized by $\Phi$. This weighting allows us to compute $\Omega_{\text{GW}}(f)$ while accounting for the desired population properties. 

% \section{O3 Full Parameters Results }
% \label{Appendix B: O3 corner}
% \begin{figure*}[h]
% \centering
% \includegraphics[width=\textwidth]{Figures/corner_allparams_O3.png}
% \caption{As Fig. \ref{fig:corner cosmo}, but using only the O3 real data. }
% \label{fig: corner all O3}
% \end{figure*}
% This section present the complete parameter estimation from the first three observational run. We adopted the Multipeak mass model (MLP), which is the most favored in recent literature, along with different priors for these parameters, listed in Tab \ref{tab:priors}. Figure \ref{fig: corner all O3} includes posterior distributions for all hyperparameters inferred from the data, offering a comprehensive view of the constraints on resolved CBC and the SGWB. 

% \section{SGWB Only Full Parameters Results}
% \label{appenix C: sgwb only}
% \begin{figure*}[h]
% \centering
% \includegraphics[width=\textwidth]{Figures/corner_stoch_only_2yr.png}
% \caption{As Fig. \ref{fig:corner cosmo}, but using only the SGWB simulated data. }
% \label{fig: corner all stoch only}
% \end{figure*}

% This section present the complete parameter estimation for the SGWB using simulated data. These results include posterior distributions for all the hyperparameters $\Phi$. The corner plot in Fig. \ref{fig: corner all stoch only} illustrate parameter correlations, helping to assess the effectiveness of the inference and the impact of unresolved sources.

\end{appendix}

\bibliographystyle{aa}
\bibliography{main}

\end{document}